\documentclass[a4paper]{article}

\newtheorem{theo}{Theorem}

\newtheorem{lemma}[theo]{Lemma}

\usepackage{type1cm}   
\usepackage{dsfont}         
\usepackage{makeidx}         
\usepackage{graphicx}        
\usepackage{multicol}        
\usepackage[bottom]{footmisc}

\usepackage{newtxtext}       %
\usepackage{newtxmath}       
\usepackage{pict2e}

\usepackage{algorithm}
\usepackage{algpseudocode}
\algtext*{EndIf}
\algtext*{EndFor}
\usepackage{amsmath}
\usepackage{subcaption}
\usepackage{mathrsfs}
\usepackage{tikz}
\usepackage{mathbbol}

\newcommand{\wabs}[1]{{\left| {#1} \right|}}

\newcommand{\wcal}[1]{{\cal {#1}}}

\newcommand{\wceil}[1]{\left\lceil {{#1}} \right\rceil }

\newcommand{\wfc}[2]{{#1}\!\left(#2\right)}

\newcommand{\wfloor}[1]{\left\lfloor {{#1}} \right\rfloor }

\newcommand{\whs}[1]{\hspace{#1cm}}

\newcommand{\wn}{{\mathds N}}

\newcommand{\wqed}{\hfill \ensuremath{\blacksquare}}
\newcommand{\wqes}{\hfill \ensuremath{\blacktriangle}}

\newcommand{\wlr}[1]{\left({#1}\right)}

\newcommand{\wo}[1]{\overline{#1}}

\newcommand{\wref}[1]{$\wlr{\ref{#1}}$}

\newcommand{\wrm}[1]{\mathrm{#1}}

\newcommand{\ws}[1]{\wcal{#1}}
\newcommand{\wset}[1]{\left\{ {#1} \right\}}
\newcommand{\wtr}[1]{\mathrm{T}}
\newcommand{\wu}[1]{\underline{#1}}

\newcommand{\wup}[1]{\wfc{\wrm{up}}{#1}}
\newcommand{\wfma}[1]{\wfc{\wrm{fma}}{#1}}

\usepackage{authblk}
\title{Computing the exact sign of sums of products with floating point arithmetic}
\author[1]{Walter F. Mascarenhas}
\affil[1]{Departamento de Computação, IME\\ Universidade de São Paulo, Brazil}
\date{\vspace{-5ex}}
\begin{document}
\maketitle

\begin{abstract}
In computational geometry, the construction of essential
primitives like convex hulls, Voronoi diagrams and Delaunay
triangulations require the evaluation of the signs of determinants,
which are sums of products. The same signs are needed for the exact 
solution of linear programming problems and systems of
linear inequalities.  Computing these signs exactly 
with inexact floating point arithmetic is challenging, and we present 
yet another algorithm for this task. Our
algorithm is efficient and uses only of floating point arithmetic,
which is much faster than exact arithmetic. We prove that the
algorithm is correct and provide efficient and tested
\texttt{C++} code for it.
\end{abstract}

\section{Introduction}
\label{sec_intro}

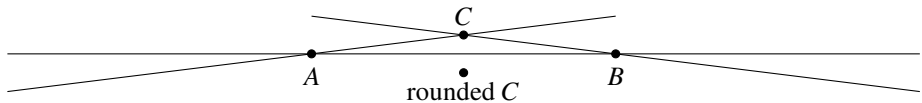
\begin{figure}[!h]
\begin{tikzpicture}[scale = 1]

\draw (0,0)--(12,0);
\draw (0,-0.5)--(8,0.5);
\draw (12,-0.5)--(4,0.5);

\filldraw[black] (4,0) circle(1.5pt);
\filldraw[black] (6,0.25) circle(1.5pt);
\filldraw[black] (8,0) circle(1.5pt);
\draw node at (6,0.5){$C$};
\draw node at (4,-0.3){$A$};
\draw node at (8,-0.3){$B$};
\draw node at (6,-0.5){rounded $C$};
\filldraw[black] (6,-0.25) circle(1.5pt);

\end{tikzpicture}
\caption{Rounding errors may lead to the conclusion that
the point $C$ lies in the wrong side of the line $AB$.}
\label{fig1}
\end{figure}

From a computational geometry perspective, we address the following
problem:  given a point $C$, defined explicitly or implicitly as the intersection
of two lines, we want to decide on which side of a line
$AB$ it lies using only floating point arithmetic
(or to find whether it is on the line.) This problem is classic.
It plays a fundamental role in the computation of convex hulls,
Voronoi diagrams and Delaunay triangulations \cite{CG,Ratschek,Shewchuk}.
The essence of the geometric issues is described in Figure \ref{fig1}. 

From an algebraic perspective, the determination of the position
of the point with respect to the line is equivalent to the computation
of the sign of a sum of products 
\[
S = \sum_{i= 1}^n \prod_{j = 1}^{n_i} a_{ij}.
\]
The inexactness of floating point arithmetic makes 
it hard to compute this sign exactly in some cases. There
are already several references about the computation
sums of floating point numbers 
\cite{Demmel,Espelid,Graillat,Higham,Kahan,Lange,WM2,Ozaki,Priest,Rump,Shewchuk,Zhu},
with applications in computations geometry.
The idea of decomposing products in sums is also
as old as \cite{Dekker,Ogita},
and the fourth chapter of \cite{Ratschek} presents an algorithm
for obtaining the signs of such sums.
However, all theses references differ in one way or another
from the present article. In summary, none of them gives
a complete solution to the problem we solve here, using
IEEE floating point arithmetic and taking underflow into account.
We present complete algorithm for this task, 
in gory detail. In particular, we take as input the factors $a_{ij}$ of the
products $p_i$ and before applying most of the other algorithms
we would need to obtain an exact representation of the 
products $p_i$, or assume that such a product does not underflow. 
The algorithm was implemented in \texttt{C++}, and it was carefully tested,
and is supported by a detailed theory. It is quite likely
that the theoretical results presented here can be generalized to bases
other than two and other rounding modes. We are not interested
in such generalizations. The goal of our theory is only
to provide a justification to what matter most us: the practical
algorithms we present, and our theoretical results are only meant
to justify our algorithms.

Finite floating point numbers suffice to solve our problem in the
situations usually found in practice,
and we only care about such numbers, 
which are elements of a set $\ws{F}$.
Deviating from the tradition, we assume that 
the arithmetic operations $\wrm{op} \in \wset{+,-,*}$
are performed rounding up. For instance, we assume 
that the subtraction of floating point numbers is defined as
\[
b \ominus a = \wup{b - a}
\]
where $\wrm{up}$ is the function from $\wlr{\min \ws{F}, max \ws{F}} \rightarrow \ws{F}$ defined by
\begin{equation}
\label{up}
\wup{x} := \min \wset{y \in \ws{F} \ \wrm{with} \ y \geq x},
\end{equation}
and the operations $\oplus$ and $\otimes$ are similar
(we have no need for expensive $\oslash$'s.)
For brevity, we leave the result of the arithmetic operation
$a \ \wrm{op} \ b$ undefined when
\[
a \ \wrm{op} \ b \not \in \wlr{\min \ws{F}, max \ws{F}}.
\]

In practice, when the rounding mode is not upwards already,
enforcing our assumption requires a function
call to change the rounding mode upwards in the beginning
of the use our functions, and another
function call to restore it when we are done, to be polite.
The cost of this two calls is amortized, but cases like
this make us believe that always rounding upwards (or always
rounding downards) is a good option for code requiring
exact results. Our Moore library \cite{Moore} is a good
example of this. Since it works with rounding mode
upwards by default, there would be no need for
changes in the rounding mode when executing the algorithms
described here.

The motivation for this article is to have a bullet proof
algorithm for computing the sign of 
$S = \sum_{i = 1}^n \prod_{j =1}^{n_i} a_{ij}$
with floating point arithmetic. Such an algorithm is needed
because sometimes rounding errors may lead to wrong conclusions by
naive algorithms. However, floating point arithmetic is an excellent
tool: such naive algorithms will be correct most of the time,
and we should care about both precision and performance.
For this reason, we believe that we should proceed in two steps:
we first try to compute the sign using a quick algorithm, which
is unable to compute the sign only in rare cases. In such rare cases
we resort to a robust algorithm, which can handle all cases but is
more expensive. We present both algorithms below.
In Section \ref{sec_quick} we describe a simple algorithm,
which is quite efficient but may not find the sign in rare
cases. In Section \ref{sec_robust} we present a robust
algorithm, which is more expensive and should be used
only after a quick algorithm was unable to find the sign.

Regarding the efficiency of our algorithms, we emphasize that
the quick one will suffice in the overwhelming majority of the
cases found in practice, and the robust algorithm will be
an extreme safety measure. As a result, usually 
the cost of evaluating the
sign will be twice the cost of evaluating the sum using
naive floating point arithmetic, plus the cost of
$\sum_{i = 1}^n n_i - n$ branches, plus the change of
rounding modes when they are necessary. This is a
$\wfc{O}{\sum_{i=1}^n n_i}$ cost, with a small constant.
However, in rare cases in which the robust algorithm
is necessary, the cost can grow exponentially
with the $n_i$.

Finally, the actual code
is implemented using the template features of \texttt{C++}, and exploiting
the details of this language it is possible to generate code
that will only resort to more expensive operations in the rare
situations in which they are needed.

\section{Quick sign}
\label{sec_quick}
In this section we present a fast algorithm which finds 
the sign of the sum of products $S := \sum_{i=1}^n \prod_{j=1}^{n_i} a_{ij}$ 
in most cases, but which may be inconclusive sometimes.
The algorithm returns a sign $s \in \wset{-1,0,1}$. If $s \neq 0$
then it definitely is the sign of the sum. However, if $s = 0$
then the sign can be anything. In this case we must resort
to a more expensive algorithm to find the sign, like the one in the next section.

As in the rest of this article, we use floating point arithmetic rounding upwards,
with at most two changes in rounding mode. 
In essence, for each product $p_i = \prod_{j=1}^{n_i} a_{ij}$ the algorithm computes 
numbers $d_i$ and $u_i$ such that 
\[
-d_i \leq p \leq u_i.
\]
If $\sum d_i < 0$ then $S$ is positive and if $\sum u_i < 0$ then $S$ is negative, otherwise we cannot decide, and the algorithm returns $0$.
The algorithm is described as Algorithm \ref{alg:quick} below,
which uses the auxiliary Algorithm \ref{alg:quick_prod}.

\begin{algorithm}[!h]
\caption{Auxiliary Algorithm for bounding aij * x...}
\label{alg:quick_prod}
\begin{algorithmic}
\Procedure{quick\_prod}{d, u, aij, x...}  

\If{ aij $\leq$ 0 }  \Comment{$-d \leq \prod_{k = 0}^{j-1} a_{ik} \leq u$}
	\If{aij $=$ 0}
	  \State \Return 0,0
	\Else                  \Comment{$-u \leq -\prod_{k = 0}^{j-1} a_{ik} \leq d$}
	 \State \Return quick\_prod(u, d, -aij, x...)  \Comment{$a_{ij} \prod_{k = 0}^{j-1} a_{ik} = -a_{ij} * -\prod_{k = 0}^{j-1} a_{ik}$}
	\EndIf

\EndIf

\State d $\gets$ d * aij \Comment{$-\wlr{d \otimes a_{ij}} \leq -\wlr{d a_{ij}} \leq \prod_{k = 0}^j a_{ik}$}
\State u $\gets$ u * aij \Comment{$\prod_{k = 0}^j a_{ik}\leq u a_{ij} \leq u \otimes a_{ij}$}

\If{is\_empty(x)}
	\State \Return d, u
\Else
	\State \Return quick\_prod(d, u, x...)
\EndIf
\EndProcedure
\end{algorithmic}
\end{algorithm}

\begin{algorithm}[!h]
\caption{A quick algorithm for computing the sign of Sum p...}
\label{alg:quick}
\begin{algorithmic}
\Procedure{quick\_sign}{p...}  
	\State round(up) \Comment{round object, resets the rounding mode on exit}
	
	\State su $\gets$ 0 \Comment{ su is an upper bound on $S$}	
	\State sd $\gets$ 0 \Comment{ sd is an upper bound on $-S$}.
	
 \For{\! {\bf each} p}
 	 \State d, u $\gets$ quick\_prod(-1, 1, p)
 	 \State su $\gets$ su + u
	 \State sd $\gets$ sd + d
 \EndFor

 \State \Return sd $<0$ ? 1 : (su $<$ 0 ? -1: 0)
	
\EndProcedure
\end{algorithmic}
\end{algorithm}

\section{Robust sign}
\label{sec_robust}

This section describes Algorithm \ref{alg:sum}, 
which computes the
sign of $S = \sum_{i =1}^n \prod_{j=1}^{n_i} a_{ij}$, using binary floating point arithmetics which
have subnormal numbers. 
The most relevant arithmetics
in this class are the ones covered by he IEEE 754 and IEEE 854 standards.
The algorithm is presented in the last page of the article. 
It assumes that there is not overflow in 
the products $p_i = \sum_{j=1}^{n_i} a_{ij}$, but
underflow is handled correctly. 
It also assumes that 
\begin{equation}
\label{boundni}
\sum_{i \ \wrm{with} \ n_i = 1} 1 + 
\sum_{i \ \wrm{with} \ n_i > 1} 2^{n_i - 2} < 1/\epsilon,
\end{equation}
where $\epsilon$ is the machine precision.
In practice, the largest $\epsilon$ we care about
corresponds to the type \texttt{float}. In this case $1/\epsilon = 2^{23}$ 
and the algorithm could be used to compute the signs of determinants
of dimensions up to 8, because if $n = 8!$ and 
$n_i = 8$ then
\[
\sum_{i=1}^n 2^{n_i - 2} = 8! \times 2^6 = 2580480 < 8388608 = 2^{23}.
\]                                     
Since $n_i = 8$ is already more than enough for the usual 
applications in computational geometry, we have no motive
to make the algorithm more complicated than it already 
is in order to relax the condition \wref{boundni}.

The algorithm is based upon two lemmas. 
Lemma  \ref{lem_sub} is about the exact computation of
the difference $b - a$  of floating point numbers.
There are versions of this lemma since the late 1960s
\cite{Dekker,Kahan,Lange}, but
we prove it here for the case in which we round upwards for completeness,
and because the details are not obvious (as stated, Lemma \ref{lem_sub}
is false if we round downwards for instance.)
In essence, it states that we can represent $b - a$ exactly as the difference
of two floating point numbers $c$ and $e$, with the
additional feature that $e$ is much smaller than $c$. As a result,
in most cases we can base our decision regarding signs on
$c$, and $e$ is used only in the rare cases in which
knowing $c$ is not sufficient.

\begin{lemma}
\label{lem_sub}
if $a,b \in \ws{F}$,  $0 < a < b$ and
\[
c := b \ominus a, \whs{1} d := b \ominus c \whs{1} 
\wrm{and} \whs{1} e := a \ominus d
\]
then
\begin{equation}
\label{thesis_sub}
b - a =  c - e \whs{1} \wrm{and} \whs{1} 0 \leq e < c \epsilon,
\end{equation}
where $\epsilon$ is the machine precision. \wqes{}
\end{lemma}

Lemma \ref{lem_fma} is the analogous to Lemma \ref{lem_sub} for multiplication,
but it is more subtle. It relies on the fused multiply add operation
($\wrm{fma}$), which is available in most processors and programming
language these days. In other words, we assume that given
$a,x, y \in \ws{F}$ such that $ax + y \in \wlr{\min \ws{F}, \max \ws{F}}$
we can compute
\begin{equation}
\label{fma}
\wfma{a,x,y} := \wup{a x + y}.
\end{equation}
It is well known that using the fma we can
represent the product $a b$ as the difference 
of two floating point numbers,
but we are not aware of proofs (or even
statements) of results describing conditions under which
this representation is exact when rounding upwards. It is important to notice that
such conditions are necessary, because Lemma \ref{lem_fma}
may not hold if the condition \wref{hypo_fma} is violated.
In order to state the decomposition result for multiplications
we need to define some constants that characterize the floating
point arithmetic:

\begin{itemize}
\item $\nu$ is the smallest positive normal element of $\ws{F}$
and $\nu \epsilon$ is the smallest positive element of $\ws{F}$
\item $\epsilon \in \ws{F}$ is the machine precision, that is, $1 + \epsilon$
is the successor of $1$ in $\ws{F}$.
\item $\sigma$ is the largest power of two in $\ws{F}$, and
we assume that $1/\sigma\in \ws{F}$.
\end{itemize}
Using the constants above we define the threshold
\begin{equation}
\label{tau}
\tau := 2 \nu / \epsilon.
\end{equation}
The values of these constants for the usual arithmetics are
presented in Table \ref{tab1}. By inspecting this table, readers will
notice that the following assumptions used in 
Lemma \ref{lem_fma} are satisfied:
\begin{equation}
\label{asa}
\sigma \epsilon^{2} \geq 2 \whs{1} \wrm{and} \whs{1}
\sigma^2 \nu \epsilon  > 2.
\end{equation}

\begin{table}[h!]
\centering
 \begin{tabular}{c | c | c | c | c} 
     & \texttt{float} & \texttt{double} & \texttt{long double}   & \texttt{quad} \\ 
\hline
$\nu$          & 1.2e-38 & 2.2e-308 & 3.4e-4932 & 3.3e-4932 \\ [0.5ex] 
$\epsilon$     & 1.2e-07 & 2.2e-016 & 1.1e-0019 & 1.9e-0034 \\ [0.5ex] 
$\sigma$       & 1.7e+38 & 9.0e+307 & 5.9e+4931 & 5.9e+4931 \\ [0.5ex] 
$\tau$         & 2.0e-31 & 2.0e-292 & 6.2e-4913 & 3.4e-4898 \\ [0.5ex] 
\end{tabular}
\caption{Values of the constants for the usual arithmetics.}
\label{tab1}
\end{table}

We now state the decomposition lemma for multiplications.
\begin{lemma}
\label{lem_fma}
Consider $a,b \in \ws{F}$, with $0 < a \leq b$, 
for which $a \otimes b$ is defined. If 
\begin{equation}
\label{hypo_fma}
c := a \otimes b > \tau,
\end{equation}
then 
\begin{equation}
\label{thesis_fma}
a b = c - d \whs{1} \wrm{for} \whs{1}  d := \wfc{\wrm{fma}}{- a, b, c} \geq 0.
\end{equation}
If the condition \wref{hypo_fma} is not satisfied then
$\tilde{a} := \sigma  \otimes a = \sigma  a \in \ws{F}$ and if
\begin{equation}
\label{hypo_fma2}
\tilde{c} := \tilde{a} \otimes b  > \tau,
\end{equation}
then
\begin{equation}
\label{thesis_fma2}
a b = \sigma^{-1} \tilde{c} - \sigma^{-1} \tilde{d} \whs{1}
\wrm{for} \whs{1}  \tilde{d} := \wfc{\wrm{fma}}{- \tilde{a}, b, \tilde{c}} \geq 0.
\end{equation}
Finally, if Equations \wref{hypo_fma} and \wref{hypo_fma2} are not satisfied 
and the first inequality in Equation \wref{asa} holds then 
 $\hat{b} := \sigma \otimes b = \sigma b \in \ws{F}$,
and if the second inequality in Equation \wref{asa} holds then
\begin{equation}
\label{thesis_fma3}
a b = \sigma^{-2} \hat{c} - \sigma^{-2} \hat{d}
\whs{0.5} \wrm{for} \whs{0.5} 
\hat{c} := \tilde{a} \otimes \hat{b} \whs{0.5} \wrm{and} \whs{0.5} 
\hat{d} := \wfc{\wrm{fma}}{ - \tilde{a}, \hat{b},\hat{c}} \geq 0.
\end{equation}
In summary, if $0 < a \leq b$ and $a \otimes b$ is defined then
there exists $e \in \wset{0,1,2}$ and $c,d \in \ws{F}$ such that 
\begin{equation}
\label{thesis_fma4}
a b = \sigma^{-e} c - \sigma^{-e} d.
\end{equation}
\wqes{}
\end{lemma}

In words, Lemma \ref{lem_fma} shows that we may fail twice in trying
to represent exactly the product $ab$ as a difference of two floating point numbers,
but the third time is a charm: we finally can 
represent $ab$ exactly as the scaled difference of two floating point
numbers. Scaling is essential here in order to deal with underflow.
We scale numbers by multiplying them by the constant $\sigma$, which
is a power of two. Such scaling does not introduce rounding errors,
but requires some book keeping. In \texttt{C++} we can keep the books using an struct like
\begin{verbatim}
struct scaled_number {
  scaled_number(T t, int exp)
  T t; 
  int exp;
};
\end{verbatim}
where $\texttt{T}$ is the type of the floating point numbers.
An scaled number $\texttt{s}$ represents 
\[
x = \wfc{\wrm{value}}{\texttt{s}} 
= \sigma^{-\texttt{s.exp}} \ \texttt{s.t}.
\]
We keep the scaled numbers in two heaps, one for the positive
values, called $\ws{P}$,  and another for the negative values,
called $\ws{N}$. In the $\ws{N}$ heap we store the absolute value
of the corresponding numbers,
so that the \texttt{t} field in our scaled numbers is always positive,
and $\texttt{exp} \leq 0$.
The elements in the heaps are sorted in increasing order
according to the following comparison function:
\begin{verbatim}
bool is_less(scaled_number x, scaled_number y) {
  if( x.exp > y.exp ) return true;
  if( x.exp < y.exp ) return false;	
  return x.t < y.t;
}  
\end{verbatim}
In order to ensure the consistency of the order above
we only push two kinds of scaled numbers in our heaps:
\begin{eqnarray}
\label{scale0}
\wrm{Scaled \  numbers} \ \texttt{s} \  \wrm{with} & & 
\texttt{s.exp} = 0 \ \wrm{and} \ \tau < \texttt{s.t},\\
\label{scale1}
\wrm{Scaled \ numbers} \ \texttt{s} \ \wrm{with} & & \ \texttt{s.exp} > 0
\ \wrm{and} \ \tau < \texttt{s.t} \leq \sigma \tau.
\end{eqnarray}
We assume that these conditions are enforced by the constructor
of \texttt{scaled\_number}s, which is only called with
positive $t$'s. We then have the following Lemma

\begin{lemma}
\label{lem_order} Under the conditions \wref{scale0} and \wref{scale1} for 
scaled numbers $\texttt{x}$ and $\texttt{y}$ we have
\[
\wfc{\wrm{value}}{\texttt{x}} < \wfc{\wrm{value}}{\texttt{y}} 
 \ \Leftrightarrow \ \wfc{\texttt{is\_less}}{\texttt{x},\texttt{y}}. 
\]
\wqes{}
\end{lemma}

We have now all the ingredients to describe our algorithm.
It uses an auxiliary function $\texttt{split}$ which splits each product
$p_i = \prod_{j=1}^{n_i} a_{ij}$ as a sum of $2^{n_i - 1}$ 
scaled numbers using Lemma \ref{lem_fma} (given this lemma, 
writing such a function is trivial.) If $n_i > 1$ then
half of the parts in which $p_i$ is split 
will be negative and the other half will be positive. Therefore,
if $n_i > 1$ then $p_i$ contributes $2^{n_i - 2}$ scaled numbers to each heap.
As a result, the left hand side of Equation \wref{boundni} is the maximum
number of elements which we will have on each heap,
and the condition \wref{boundni} ensures that 
this number does not exceed $1/\epsilon$. 

Once the heaps are filled with products, we start to compare them.
While
\[
s_n := \wfc{\wrm{size}}{\ws{N}} > 0 \whs{1} \wrm{and} \whs{1}
s_p := \wfc{\wrm{size}}{\ws{P}} > 0
\]
we pop the top elements $n$ and $p$ of $\ws{N}$ and $\ws{P}$ 
and compare them. If $p > s_n * n$ then the sign of the sum is $1$. 
If $n > s_p * p$ then the sign of the sum is $-1$, 
otherwise, conceptually, we use Lemma \ref{lem_sub} to split
$p - n$ and push the parts back into the heaps. There is a catch
in this argument in that $n$ and $p$ are scaled numbers, which
may have different exponents. If these exponents differ by
more than one then the numbers on the heap with the numbers with
the largest exponent are negligible and we are done. When 
the exponents differ by one
we multiply the \texttt{t} field of the one with the largest
exponent by $\sigma$, reducing both numbers to the same exponent.
This multiplication by $1/\sigma$ may be inexact, but this
inexactness is harmless. For instance, when $p$ has the 
largest exponent and the multiplication $z := p.t \gets p.t \otimes 1/\sigma$
is inexact, then Lemma \ref{lem_twice} in Section \ref{sec_proofs}
implies that $z \leq \nu$ and Equations \wref{scale0}
and \wref{scale1} yield
\begin{equation}
\label{boundni2}
n.t > \tau =  \frac{2}{\epsilon} \nu \geq \frac{2}{\epsilon} z \geq 2 s_p z,
\end{equation}
and we will reach the correct conclusion that the sign
of the sum is $-1$ even
if we use the incorrect $z$.  Once we have both numbers 
with the same exponent, simply split the difference
$p.t - n.t$ and adjust the exponents of the results consistently.

Finally, the algorithm terminates because we have two possibilities
after we reduce $p$ and $n$ to the same exponent.
When $n.t \leq p.t$ (and the case $n.t > p.t$ is analogous.):
\begin{itemize}
\item[(i)] If $p.t \ominus p.n < p.t$ then the largest $t$ field decreases,
and this can only happen a finite number of times.
\item[(ii)] If $p.t \ominus p.n = p.t$ then Lemma \ref{lem_sub} implies
that $e = p.n < \epsilon p.t$. Since the number of elements
in $\ws{N}$ is at most $1/\epsilon$ by bound \wref{boundni}, this
implies that $p.t > s_n * p.n$, and the algorithm returns $1$
due to this condition.
\end{itemize}

This is only a high level description of the algorithm.
A reasonably detailed version of it is presented in the
last page of this article. The actual code is a bit
more involved, due to optimizations which replace
scaled\_numbers by plain floating point numbers when
possible. Readers interested in the 
implementation details should look at the \texttt{C++} code available
as supplementary material to the arxiv version of this article. 
This code is distributed under the Mozilla Public License 2.0.

\begin{algorithm}[!h]
\caption{Sign of Sums of Products}
\label{alg:sum}
\begin{algorithmic}[1]
\Procedure{sign}{p...} \Comment{sign of the sum p[0] + p[1] +...}

 \State round r(up)      \Comment{Rounding object, resets the mode on exit}
 \State heap neg, pos    
 \State
 \For{\! {\bf each} p }  
 	 \State split(neg, pos, p) \Comment{Each p is a product} 
 \EndFor
  \State

 \While{ true }
   \State sn, sp $\gets$ size(neg), size(pos)
   \State
   \If{ sn $=$ 0 } 
		 \State \Return (sp $=$ 0) ? 0 : 1 
   \EndIf 
 
   \If{ sp $=$  0 } 
     \State \Return -1 
  \EndIf
	\State
   \State n, p $\gets$ pop(neg), \ pop(pos)

  \State
   \If{ n.exp $\leq$ p.exp} \Comment{small n.exp $\Rightarrow$ large n $=2^{-\wrm{n.exp}}$ n.t }
   \If{ n.exp $<$ p.exp }  
     \If{ n.exp $<$ p.exp - 1} 
       \State \Return -1 \Comment{p is too small} 
     \EndIf
     \State p.t $\gets$ p.t * $1/\sigma$  \Comment{$p.t / \sigma$ is exact or -1 is returned in line \ref{tiny_ret}}
  \EndIf
  
  \If{ n.t $>$ sp * p.t } 
     \State \Return -1  \label{tiny_ret}
  \EndIf
   \State 
  \If{ n.t $ > $ p.t }                         
		 \State c, e $\gets$ lemma\_1(p.t, n.t) \Comment{split n.t - p.t as in Lemma \ref{lem_sub}}
		 \State neg $\gets$ scaled\_number(c, n.exp)
		 \If{e $>$ 0} 
		    \State pos $\gets$ scaled\_number(e, n.exp)  
		 \EndIf
  \Else
     \If{ p.t $ > $ n.t }
		 \State c, e $\gets$ lemma\_1(n.t, p.t) \Comment{split p.t - n.t as in Lemma \ref{lem_sub}}
		 \State pos $\gets$ scaled\_number(c, n.exp)
		 \If{e $>$ 0} 
		   \State neg $\gets$ scaled\_number(e, n.exp) 
		\EndIf
		\EndIf
	\EndIf	 
\Else     
	 \If{ n.exp $\geq$ p.exp $+$ 1 }  
	   \State \Return 1   \Comment{n is too small}
   \EndIf
	 
   \State  n.t $\gets$ n.t * 1/$\sigma$ \Comment{$n.t/\sigma$ is exact or 1 is returned in line \ref{tiny_retB}}

  \If{p.t $>$ sn * n.t} 
  	\State \Return 1  \label{tiny_retB}
  \EndIf
 	 \State c, e $\gets$ lemma\_1(n.t, p.t) \Comment{split p.t - n.t as in Lemma \ref{lem_sub}}
   \State pos $\gets$ scaled\_number(c, p.exp)
	 \If{e $>$ 0}
		 \State neg $\gets$ scaled\_number(e, p.exp)
 \EndIf
\EndIf
  
 \EndWhile

\EndProcedure
\end{algorithmic}
\end{algorithm}

\section{Proofs}
\label{sec_proofs}
Here we prove the results stated above and two auxiliary results.
Our proofs use the following characteristics
shared by the usual binary floating point arithmetics with
subnormal numbers, like the ones covered by the IEEE
standards 754 and 854. There are three kinds of elements
in the set $\ws{F}$ of finite floating point numbers:
\begin{itemize}
\item $0$ is a floating point number.
\item $x \in \ws{F}$ if and only if $-x \in \ws{F}$.
\item The subnormal numbers $x$ have absolute value of the form
\begin{equation}
\label{subnormal}
x = \nu \epsilon m, \whs{1} \wrm{with} \whs{1} m \in 
\wn{} \whs{1} \wrm{and} \whs{1}
1 \leq m < 1/\epsilon.
\end{equation}
\item The normal numbers $x$ have absolute value of the form
\begin{equation}
\label{normal}
x = 2^{e} \nu \epsilon m
\end{equation}
for integers $e$ and $m$ such that
\begin{equation}
\label{normal2}
0 \leq e \leq e_{\max} := \wfc{\log_2}{\sigma} - \wfc{\log_2}{\nu}
\whs{1} \wrm{and} \whs{1}
1 \leq \epsilon m < 2.
\end{equation}
\end{itemize} 

We use two auxiliary lemmas, and the proofs are presented after
the statements of these lemmas. The lemmas are proved in
the order in which they were stated.

\begin{lemma}
\label{lem_twice}
If $x \in \ws{F}$ and $2 \wabs{x} \leq \max \ws{F}$ then $2 x \in \ws{F}$. Therefore, if $e$ is a positive integer and  $2^e \wabs{x} \leq \max \ws{F}$ then $2^e x \in \ws{F}$.
Similary, if $k\geq 0$ and $2^{-k} \wabs{x} \geq \nu$ then 
$2^{-k} \otimes x = 2^{-k} x$ is normal.
\wqes{}
\end{lemma}

\begin{lemma}
\label{lem_r2}
If the integer $\ell \geq 0$ and the real number $x$ are such that
\begin{equation}
\label{hypo_r2p}
2^{\ell} \leq \frac{x}{\nu} \leq 2^{\ell + 1} \leq \sigma
\end{equation}
then 
\begin{equation}
\label{thesis_r2p}
\wup{x} = 2^{\ell} \nu \epsilon \wceil{\frac{x}{2^{\ell} \nu \epsilon}}.
\end{equation}
\wqes{}
\end{lemma}


{\bf Proof of Lemma \ref{lem_sub}.}
If $b = \nu \epsilon m_b$ is either subnormal or normal
with a minimum exponent, then $a$ is also of the form
$a = \nu \epsilon m_a$,
\[
c = \nu \epsilon \wlr{m_b - m_a}, \whs{1} d = \nu \epsilon m_a, \whs{1} \
e = 0 \whs{1} \wrm{and} \whs{1} b - a = c = c + e,
\]
and Equation \wref{thesis_sub} holds.
We can then assume that 
\begin{equation}
\label{bsub}
b = 2^{e_b} \nu \epsilon m_b \whs{1} \wrm{with} \whs{1}
e_b > 0 \whs{1} \wrm{and} \whs{1} 1 \leq \epsilon m_b < 2,
\end{equation}
and
\[
a = 2^{e_a} \nu \epsilon m_a \whs{1} \wrm{with} \whs{1}
e_a \geq 0 \whs{1} \wrm{and} \whs{1} 1 \leq m_a < 2 / \epsilon.
\]
It follows that
\begin{equation}
\label{subh}
b - a = \nu \epsilon h \whs{1} \wrm{for} \whs{1}
h := 2^{e_b} m_b - 2^{e_a} m_a > 0.
\end{equation}
If $h < 1/\epsilon$ then $b - a$ is subnormal and
\[
c = b \ominus a = b - a \ \Rightarrow \ d = a \ \Rightarrow \ e = 0
\ \Rightarrow \ b - a = c - e,
\]
Equation \wref{thesis_sub} holds and we are done. 
Let us then assume that $h \geq \epsilon$
and let $\ell \geq 0$ be the integer such that
\begin{equation}
\label{squeeze_sub}
2^{\ell} \leq \frac{b - a}{\nu} = \epsilon h \leq 2^{\ell + 1}.
\end{equation}
By Lemma \ref{lem_r2},
\begin{equation}
\label{csub}
c = b \ominus a = 2^{\ell} \nu \epsilon \wceil{2^{-\ell} h}
= 2^{\ell} \nu \epsilon \wo{c} 
\end{equation}
for
\begin{equation}
\label{cbar}
\wo{c} := \wceil{2^{e_b -\ell} m_b - 2^{e_a -\ell} m_a}.
\end{equation}
Equation \wref{squeeze_sub} leads to the bound 
\begin{equation}
\label{hsub}
h = 2^{e_b} m_b - 2^{e_a} m_a \geq 2^{\ell} / \epsilon
\end{equation}
and Equation \wref{hsub} shows that
\[
2^{e_b} m_b \geq 2^{\ell}/\epsilon \ \Rightarrow \ 
2^{e_b} \geq \frac{2^{\ell}}{\epsilon m_b} > 2^{\ell - 1}
\Rightarrow e_b \geq \ell,
\]
and $2^{e_b - \ell} m_b$ is integer. 

If $2^{e_a -\ell} m_a < 1$ then 
\[
\wo{c} = 2^{e_b - \ell} m_b, \whs{0.5} c = b, \whs{0.5} d = b \ominus c = 0, 
\whs{0.5} e = a \ominus d = a \ \ \Rightarrow \ \ b - a = c - e,
\]
and the first part of Equation \wref{thesis_sub} holds.
Moreover, Equation \wref{hsub} yields
\[
e = a = 2^{e_a} m_a \nu \epsilon < 2^{\ell} \nu \epsilon < 
\wlr{2^{e_b} m_b \epsilon} \nu \epsilon = \epsilon b = \epsilon c,
\]
and the second part of Equation \wref{thesis_sub} also holds.
Therefore, we can assume that 
\begin{equation}
\label{bma}
2^{\ell - e_a} \leq m_a < 2/\epsilon.
\end{equation}
If $e_a \geq \ell$ then $2^{e_a - \ell}$ is integer,
\[
\wo{c} = 2^{e_b - \ell} m_b - 2^{e_a - \ell} m_a, 
\whs{0.5} c = b - a, \whs{0.5} d = b \ominus c = a, 
\whs{0.5} e = a \ominus d = 0 \ \ \Rightarrow \ \ b - a = c - e
\]
and Equation \wref{thesis_sub} holds again. 
Therefore, we can assume that $e_a < \ell$ and
\begin{equation}
\label{qsub}
q := \ell - e_a > 0.
\end{equation}
Equation \wref{bma} shows that $2^{-q} m_a \geq 2^{e_a - \ell} \times 2^{\ell - e_a} = 1$
and the integers
\[
\wo{a} := \wfloor{2^{-q} m_a}
\whs{1} \wrm{and} \whs{1}
\wu{a} = m_a - 2^{q} \wo{a} 
\]
are such that
\begin{equation}
\label{abar}
m_a = 2^{q} \wo{a} + \wu{a},
\whs{0.5}
1 \leq \wo{a} \leq m_a/2 < 1/\epsilon
\whs{0.5} \wrm{and} \whs{0.5} 0 
\leq \wu{a} < 2^{q} \leq m_a < 2/\epsilon.
\end{equation}
It follows that
\[
2^{e_b - \ell} m_b - 2^{e_a - \ell} m_a = 
2^{e_b - \ell} m_b - 2^{-q} m_a = 
2^{e_b - \ell} m_b - \wo{a} - 2^{-q} \wu{a}
\]
and the bound $0 \leq \wu{a} < 2^{q}$ leads to
\[
\wo{c} = 
\wceil{2^{e_b - \ell} m_b - 2^{e_a - \ell} m_a}
=  2^{e_b - \ell} m_b - \wo{a}
\]
and
\begin{equation}
\label{subc}
c = b \ominus a = b - 2^{\ell} \nu \epsilon \wo{a} = b - \hat{a}
\whs{1} \wrm{for} \whs{1}
\hat{a} := 2^{\ell} \nu \epsilon \wo{a}.
\end{equation}
The second bound in Equation \wref{abar} shows that
$\nu \epsilon \wo{a}$ is subnormal and Lemma \ref{lem_twice} 
shows that $\hat{a} \in \ws{F}$. This implies that
\begin{equation}
\label{subd}
d = b \ominus c = b - c = \hat{a}.
\end{equation}
Additionally,
\begin{equation}
\label{subad}
a - d = 2^{e_a} \nu \epsilon m_a - 2^{\ell} \nu \epsilon \wo{a}
= 2^{e_a} \nu \epsilon \wlr{ m_a - 2^{q} \wo{a}}
= 2^{e_a} \nu \epsilon \wu{a} =: \tilde{a}.
\end{equation}
If $\wu{a} = 0$ then $\tilde{a} \in \ws{F}$. 
If $\wu{a} < 1/\epsilon$ then the same argument used for $\hat{a}$ shows that 
$\tilde{a} \in \ws{F}$, and if $1/\epsilon \leq \wu{a} < 2/\epsilon$ then
$\nu \epsilon \wu{a}$ is normal and Lemma \ref{lem_twice} shows that 
$\tilde{a} \in \ws{F}$. Therefore, in all cases for $\wu{a}$
in Equation \wref{abar} we have that $\tilde{a} \in \ws{F}$ and
\begin{equation}
\label{sube}
e = a \ominus d = a - d = \tilde{a}.
\end{equation}
Equations \wref{subd} and \wref{subad} show
that $\hat{a} + \tilde{a} = a$ and Equations 
\wref{subc} and \wref{sube} yield
\[
c - e = b - \hat{a} - \tilde{a} = b - a,
\]
and the first part of Equation \wref{thesis_sub} holds. 
Finally, Equations \wref{csub}, \wref{cbar}, \wref{hsub},\wref{qsub}
and \wref{abar} imply that
\[
\epsilon c \geq 2^{\ell} \nu \epsilon  = 2^{q + e_a} \nu \epsilon  
> 2^{e_a} \wu{a} \nu \epsilon = \tilde{a}.
\]
and the second part of Equation \wref{thesis_sub} holds.
\wqed{}


{\bf Proof of Lemma \ref{lem_fma}.}
Let us start with the case in which
$c = a \otimes b > \tau = 2 \nu / \epsilon$.
In this case $b > \sqrt{2 \nu/\epsilon} > \nu$ and 
$b$ is normal. Therefore,
\begin{equation}
\label{eqb}
b = 2^{e_b} m_b \whs{1} \wrm{with} \whs{1} 1 \leq \epsilon m_b < 2.
\end{equation}
$a$ can be normal or subnormal, but in both cases there exist integers
$e_a$, $\ell_a$ and $m_a$ with
\begin{equation}
\label{eqa}
a = 2^{e_a} m_a, \whs{1} 2^{\ell_a} \leq m_a < 2^{\ell_a + 1}
\whs{1} \wrm{and} \whs{1} 1 \leq 2^{\ell_a} < 1/\epsilon.
\end{equation} 
If $\ell > 0$ is the integer such that 
\begin{equation}
\label{squeeze}
2^{\ell} \leq \frac{a b}{\nu} < 2^{\ell + 1}
\end{equation}
then
\[
2^{\ell + 1} \geq \frac{a b}{\nu} > 1 / \epsilon 
\ \Rightarrow 2^{\ell} > 1 / \wlr{2 \epsilon} > 1 \ \ \Rightarrow \ell > 0,
\]
and Lemma \ref{lem_r2} and Equations \wref{eqb} and \wref{eqa} show that 
\begin{equation}
\label{sub6}
a \otimes b = 2^{\ell} \nu \epsilon 
\wceil{\frac{2^{e_a + e_b} m_a m_b}{2^{\ell} \nu \epsilon} }
= \wceil{\frac{m_a m_b}{p}}
\end{equation}
for
\begin{equation}
\label{pf}
p := 2^{\ell - e_a - e_b} \nu \epsilon.
\end{equation}
Since 
\[
\frac{a b}{2^{\ell} v \epsilon} = \frac{m_a m_b}{p},
\]
Equation \wref{squeeze} implies that 
\[
\frac{a b}{v} = \frac{2^{\ell} \epsilon m_a m_b}{p} < 2^{\ell+1}
\]
and Equations \wref{eqb} and \wref{eqa} yield
\[
p > \epsilon m_a m_b / 2 = \wlr{m_a/2} \wlr{\epsilon m_b} \geq 2^{\ell_a - 1} \geq 1/2
\Rightarrow p > 1/2.
\]
Since $p$ is a power of two, this implies that $p \geq 1$ is integer.
On the other hand, Equation \wref{squeeze} implies that
\[
\frac{2^{\ell} \epsilon m_a m_b}{p} \geq 2^{\ell}
\]
and Equations \wref{eqb} and \wref{eqa} lead to
\[
p \leq \epsilon m_a m_b < 4 / \epsilon \Rightarrow p \leq 2/\epsilon.
\]
Euclid's division algorithm yields integers $q$ and $r$ such that
\begin{equation}
\label{mambf}
m_a m_b = q p - r \whs{1} \wrm{and} \whs{1} 0 \leq r < p \leq 2/\epsilon, 
\end{equation}
and Equation \wref{sub6} shows that
\begin{equation}
\label{otimes}
c = a \otimes b = 2^{\ell} \nu \epsilon 
\wceil{q - p^{-1} r } = 2^{\ell} \nu \epsilon q =
a b + 2^{\ell} p^{-1} \nu \epsilon r.
\end{equation}
Equations \wref{eqb} and \wref{eqa} imply that
\[
m_a m_b < 2^{\ell_a + 1} 2/\epsilon < \frac{4}{\epsilon^2}
\ \Rightarrow \frac{1}{m_a m_b} > \frac{\epsilon^2}{4},
\]
the bound $a \otimes b > 2 \nu/\epsilon$ yields $a b \epsilon > 2 \nu$ and 
\[
2^{\ell} p^{-1} = \frac{2^{e_a + e_b}}{\nu \epsilon} =
 \frac{1}{\epsilon^2 m_a m_b} \frac{2^{e_a + e_b} m_a m_b \epsilon}{\nu}
 =  \frac{1}{\epsilon^2 m_a m_b} \frac{a b \epsilon}{\nu}
 >  \frac{2}{\epsilon^2 m_a m_b} > \frac{1}{2}.
\]
Since $2^{\ell} p^{-1}$ is a power of 2, this implies that
$2^{\ell} p^{-1}$ is integer, and the last inequality in Equation \wref{mambf}
and Lemma \ref{lem_twice} imply that
\[
d := 2^{\ell} p^{-1} \nu \epsilon r \in \ws{F}.
\]
Combining this with Equations \wref{otimes} we obtain that
\[
a b = 2^{\ell} \nu \epsilon q - d = c - d 
\]
and
\[
\wlr{-a} b + c = d \in \ws{F} \Rightarrow 
\wfma{-a, b, c} = d.
\]
This completes the proof of Equation \wref{thesis_fma}.

If $a \otimes b \leq \tau$ then
\[
a^2 \leq a b \leq \tau
\Rightarrow a \leq \sqrt{\frac{2 \nu}{\epsilon}} < 1
\Rightarrow \sigma a < \sigma \Rightarrow \tilde{a} = \sigma a = \sigma \otimes a \in \ws{F},
\]
and using the same argument used to prove Equation \wref{thesis_fma} we can prove Equation \wref{thesis_fma2}.

Finally, the smallest value possible for a positive floating
point number is $\nu \epsilon$, 
and if the conditions \wref{hypo_fma} and
\wref{hypo_fma2} are not satified then 
\[
\tilde{a} = \sigma a \geq \sigma \nu \epsilon.
\]
The violation of condition \wref{hypo_fma2} and the 
first inequality in Equation \wref{asa} yield
\[
b \leq \frac{2 \nu}{\epsilon \tilde{a}}
\leq  
\frac{2 \nu}{\epsilon \sigma \nu \epsilon}
\Rightarrow \sigma b \leq \frac{2}{\epsilon^{2}} \leq \sigma.
\]
This bound implies that $\hat{b} = \sigma \otimes b = \sigma b \in \ws{F}$.
We also have that $\hat{b} \geq \sigma \nu \epsilon$
and the second inequality in Equation \wref{asa} leads to
\[
\tilde{a} \hat{b} \geq \sigma^2 \nu^2 \epsilon^2 = 
\frac{\sigma^2 \nu \epsilon}{2} \frac{2 \nu}{\epsilon} > \frac{2 \nu}{\epsilon} = \tau.
\]
This condition allows us to use the same argument used to prove
the validity of Equations \wref{thesis_fma} and \wref{thesis_fma2}
to prove Equations \wref{thesis_fma3}, and this proof is complete.
\wqed{}


{\bf Proof of Lemma \ref{lem_order}.}
If $\texttt{x.exp} > \texttt{y.exp}$ then
$\texttt{x.exp} > 0$ and item (ii) implies that
$\texttt{x.t} \leq \sigma \tau$. It follows that
$\texttt{x.exp} \geq \texttt{y.exp} + 1$ and
using item (i) we derive that
\[
\wfc{\wrm{value}}{\texttt{x}} =
\sigma^{-\texttt{x.exp}} \texttt{x.t}
\leq \sigma^{-\texttt{y.exp} - 1} \ \sigma \tau
= \sigma^{-\texttt{y.exp}} \ \tau < 
 \sigma^{-\texttt{y.exp}} \ \texttt{y.t}
 = \wfc{\texttt{value}}{y},
\]
and the function $\texttt{is\_less}$ returns the correct value
in this branch. The branch $\texttt{x.exp} < \texttt{y.exp}$
is analogous. Finally, in the last branch the exponents
cancel out and we are left with the correct comparison
of the $\texttt{t}$ fields. \wqed{}


{\bf Proof of Lemma \ref{lem_twice}.} If $x = 0$ then $2 x = 0 \in \ws{F}$. 
If $x$ is normal, then  Lemma \ref{lem_twice} follows directly from Equation \wref{normal}.
If $x = \nu \epsilon m $ is subnormal then there are two possibilities: 
If $\wabs{2 m} < 1/\epsilon$ then $2 x = \nu \epsilon \wlr{2m}$ is also subnormal. 
If $\wabs{2 m} \geq 1/\epsilon$ then $2 x = \nu \epsilon \wlr{2m}$ is normal, because
$1/\epsilon \leq \wabs{m} < 2/\epsilon$. The part of the Lemma regarding
division follows directly from the definition of normal number.
\wqed{}


{\bf Proof of Lemma \ref{lem_r2}.} If $x$ satisfies the condition
\wref{hypo_r2p} then
\[
1/\epsilon \leq q:= \frac{x}{2^{\ell} \nu \epsilon} \leq 2/\epsilon.
\]
Therefore,
\[
1/\epsilon \leq \wceil{q} \leq 2/\epsilon.
\]
The number 
\[
\wo{x} := 2^{\ell} \nu \epsilon \wceil{q}
\]
belongs to $\ws{F}$ because if $\wceil{q} < 2/\epsilon$ then
it fits on definition \wref{normal} with $e = \ell$ and $m = \wceil{q}$,
and if $\wceil{q} = 2/\epsilon$ then
\[
\wo{x} = 2^{\ell + 1} \nu \epsilon \time 1/\epsilon
\]
and definition \wref{normal} holds with $e = \ell + 1$ and $m = 1/\epsilon$.
We have that 
\[
2^{\ell} \nu \epsilon q = x \leq \wo{x},
\]
and the definition of $\wup{x}$ in \wref{up} implies that 
$\wup{x}  \leq \wo{x}$. To prove that $\wup{x} = \wo{x}$, 
we now show that if
$y \in \ws{F}$ is such that 
$y \geq x$ then $y \geq \wo{x}$. In fact, if $y \geq x$
then $y \geq 2^{\ell} \nu$ and this implies that
$y$ is normal, that is,
\[
y = 2^{e_y} \nu \epsilon m_y \whs{0.5} \wrm{with} \whs{0.5}  1/\epsilon \leq m_y < 2/\epsilon
\whs{0.5} \wrm{and} \whs{0.5} 2^{e_y} \nu \epsilon m_y \geq x = 2^{\ell} \nu \epsilon q.
\]
This leads to
\[
 2^{e_y}  m_y \geq 2^{\ell} q
 \  \ \Rightarrow \ \ 2^{e_y - \ell} \geq  q / m_y > \wlr{1/\epsilon} / (2 / \epsilon) = 1/2
\Rightarrow e_y \geq \ell,
\]
and $2^{e_y - \ell} m_y$ is integer. As a result
\[
 2^{e_y}  m_y \geq 2^{\ell} q  \ \Rightarrow \  
  2^{e_y - \ell}  m_y \geq q  \ \Rightarrow \  
 2^{e_y - \ell}  m_y \geq \wceil{q} \  \Rightarrow \  
y =  2^{e_y} \nu \epsilon m_y \geq 2^{\ell} \nu \epsilon \wceil{q} = \wo{x},
\]
and we are done. \wqed{}


\begin{thebibliography}{}

\bibitem{CG}
De Berg. M., van Kreveld, M., Overmans, M.,
Schwarzkopf,O., Computational Geometry,
algorithms and applications, Springer (2008). 

\bibitem{Dekker}
Dekker, T.J., A floating-point technique for extending the available precision. 
Numer. Math. 18(3), 224 (1971). 

\bibitem{Demmel}
Demmel, J., Hida, Y.,
Fast and accurate floating point summation with application 
to computational geometry. Numer. Algorithms 37(1--4), 101--112 (2004). 

\bibitem{Espelid}
Espelid, T. O., On floating point summation. SIAM Review 37, 
603--607 (1995).

\bibitem{Graillat}
Graillat, S., Louvet, N., Applications of fast and accurate summation in 
computational geometry. Technical report, Laboratoire LP2A, University of Perpignan, 
Perpignan, France (2006).

\bibitem{Higham}
Higham, N. J., The accuracy of floating point summation. SIAM J. Sci.
Computation 14, 783--799 (1993).

\bibitem{Kahan}
Kahan, W.
Further remarks on reducing truncation errors. Commun. ACM 8(1), 40 (1965).

\bibitem{Lange}
Lange, M., Oishi, S., A note on Dekker's FastTwoSum algorithm. 
Numerische Mathematik, 145(2), 383--403 (2020). 

\bibitem{Moore} Mascarenhas, W.F., Moore: Interval Arithmetic in C++20,
In: Barreto G., Coelho R. (eds) Fuzzy Information Processing. NAFIPS 2018. 
Communications in Computer and Information Science, vol 831, pp 519--529 (2018). 

\bibitem{WM2} Mascarenhas, W.F., Floating point numbers are real numbers,
arXiv:1605.09202v2 [math.NA] (2017).

\bibitem{Ogita}
Ogita, T., Rump, S.M., Oishi, S,
Accurate sum and dot product. SIAM J. Sci. Comput. 26(6), 1955--1988 (2005).


\bibitem{Ozaki}
Ozaki K., Ogita, T.,  Oishi S.,
A robust algorithm for geometric predicate by error-free determinant
transformation, Information and Computation 216  3--13 (2012).

\bibitem{Priest}
Priest, D. M., On properties of floating point arithmetics: numerical
stability and the cost of accurate computations. Ph.D. Thesis, University
of California, Berkeley (1992).

\bibitem{Ratschek}
Ratschek,H. and Rokne, J., Geometric Computations with Interval
and New Robust Methods, Applications in Computer Graphics,
GIS and Computational Geometry. Horwood Publishing
Chichester (2003).

\bibitem{Rump}
Rump, S., Error estimation of floating-point summation and dot product. BIT
Numerical Mathematics 52(1) 201--220 (2012).

\bibitem{Shewchuk}
Shewchuk, J., Adaptive Precision Floating-Point Arithmetic and 
Fast Robust Geometric Predicates, Discrete \& Computational Geometry 18:305--363 (1997).

\bibitem{Zhu}
Zhu, Y.K., Hayes, W.B.,  Algorithm 908. ACM Trans. Math. Softw. 37(3), 1--13 (2010).

\end{thebibliography}
\end{document}